\documentclass{aa}  

\usepackage{graphicx}
\usepackage{txfonts}
\usepackage{lineno}
\begin{document}

   \title{VLBI observations of the high-redshift X-ray bright blazar SRGE J170245.3+130104}    
   \author{Yuanqi Liu \inst{1},
           Tao An \inst{1,2,3}\fnmsep\thanks{antao@shao.ac.cn}  ,        
           Shaoguang Guo\inst{1,2,3},
           Yingkang Zhang \inst{1,3},
           Ailing Wang \inst{1,2},
           Zhijun Xu \inst{1},
           Georgii Khorunzhev \inst{4},
           Yulia Sotnikova \inst{5},
           Timur Mufakharov \inst{5,6},
           Alexander Mikhailov \inst{5},
           \and
           Marat Mingaliev  \inst{5,6,7}
         }

   \institute{Shanghai Astronomical Observatory, Key Laboratory of Radio Astronomy, CAS, 80 Nandan Road, Shanghai 200030, China\\
         \and
             School of Astronomy and Space Sciences, University of Chinese Academy of Sciences, No. 19A Yuquan Road, Beijing 100049, China\\
         \and
            Key Laboratory of Radio Astronomy and Technology, Chinese Academy of Sciences, A20 Datun Road, Beijing, 100101, P. R. China \\
        \and Space Research Institute of RAS,Profsoyusnaya street, 84/32, 117997, Russia \\
        \and Special Astrophysical Observatory of RAS, Nizhny Arkhyz, 369167, Russia \\
        \and Kazan Federal University, 18 Kremlyovskaya St, Kazan 420008, Russia \\
        \and Institute of Applied Astronomy RAS, St. Petersburg 191187, Russia \\
             }

   \date{Received 30/01/2024; accepted 26/02/2024}

 
  \abstract
   {}
   {The X-ray luminous and radio-loud AGN SRGE J170245.3+130104 discovered at $z\sim 5.5$ provides unique chances to probe the SMBH growth and evolution with powerful jets in the early Universe. }
   {We present $1.35-5.1$ GHz Very Long Baseline Array (VLBA) results on the radio continuum emission and spectrum analysis for this quasar in a low flux density state. }
   {This source is unresolved at three frequencies with the total flux densities of $8.35 \pm 0.09 \ \rm mJy \ beam^{-1}$, $7.47 \pm 0.08 \ \rm mJy \ beam^{-1}$, and $6.57 \pm 0.02 \ \rm mJy \ beam^{-1}$ at 1.73 GHz, 2.26 GHz, and 4.87 GHz, respectively. Meanwhile, the brightness temperature is higher than $10^9$~K. }
  {Compared with previous radio observations with arcsec-scale resolution, nearly all the radio emission from this source concentrates in the very central milli-arcsecond (mas) scale area. We confirm this source is a bright blazar at $z>5$. This young AGN provide us the great chances to understand the first generation of strong jets in the early Universe.}

    \titlerunning{Liu et al. VLBI observations of J1702+1301}

   \maketitle

%

\section{Introduction}

The study of high-redshift active galactic nuclei (AGNs) plays a crucial role in advancing our understanding of the early Universe, particularly in exploring the formation and evolution of supermassive black holes (SMBHs). These distant AGNs provide a unique window into the nascent stages of SMBHs, shedding light on their accretion processes and immediate environment. Such studies are crucial for understanding the mechanisms that drive the early growth of these cosmic monsters. 

Currently, around 600 AGNs have been discovered at redshift higher than 5 (for example, \citealt{Fan2006, Willott2010, Yang2017, Yang2023}), which is close to the end of cosmic reionization. These discoveries are essential for understanding this crucial moment in the evolution of the Universe. 

Blazars are a subclass of AGN with jets aligned close to the line of sight, whose emission is relativistic beamed and Doppler boosted, providing favorable conditions for observing high-redshift AGN. Observations of jet emission from high-$z$ blazars enable tests of relativistic jet models and physical properties (such as jet viewing angle and Lorentz factor) under extreme conditions in the early Universe and constraints on jet power and particle acceleration. Statistical studies of the number density, distribution, and evolution of blazars can yield insights into the co-evolution of black holes and galaxies in the early Universe.
To date, PSO J030947.49+271757.31 at $z \sim 6.1$ (hereafter J0309+2717) is the highest redshift blazar discovered \citep{Belladitta2020}. Since very few blazars have been observed at $z>5$, discovering and studying more high-$z$ ($z>5$) blazars is extremely valuable for enriching our understanding of high-redshift AGN.

Our research focuses on SRGE J170245.3+130104 (hereafter J1702+1301), one of the highest-redshift and high-luminosity X-ray quasars recently discovered by the eROSITA telescope on-board the Spectrum Roentgen Gamma (SRG; \citealt{Merloni2020}) during its first half-year X-ray All-sky Survey \citep{Khorunzhev2021}. Follow-up the Burst Alert Telescope (BAT) telescope's spectroscopic observations measured the redshift of J1702+1301 to be $z=5.466\pm0.003$, and although only a limited spectral coverage has been obtained so far, the mass of its central black hole is estimated to be at least $10^8 M_\odot$ \citep{Khorunzhev2021}. J1702+1301 exhibits powerful X-ray and radio emission with a radio luminosity of $10^{27} ~ \rm W ~ Hz^{-1}$, a radio loudness of $R>1100$, a rapidly varying radio flux density, and a flat radio spectrum \citep{An2023}. All these observations are consistent with those of blazars, suggesting that J1702+1301 is most likely a high-$z$ blazar with a relativistic jet aligned to the line of sight. J1702+1301 presents an exceptional opportunity to investigate the first generation of strong jets emanating from young AGN. 

J1702+1301 stands out due to its extraordinary luminosity in X-ray and radio at $z>5$, making it a benchmark for follow-up observational campaigns. Its extreme characteristics make it an invaluable reference point for comparative studies of similar objects, enhancing our overall knowledge of high-redshift AGNs.
Our Very Long Baseline Array (VLBA) observations at 1.73 GHz, 2.26 GHz, and 4.87 GHz have revealed that nearly all radio emissions from J1702+1301 are concentrated in a central milli-arcsecond region, with a brightness temperature exceeding $10^9$ K \citep{An2023}. This concentration of emissions in a compact region suggests a highly active central core, a typical characteristic of blazars.
Moreover, our observations of J1702+1301 at these frequencies, corresponding to significantly higher rest-frame frequencies due to the source's redshift, enable a deeper understanding of the blazar's properties and behavior in the early universe. By comparing J1702+1301's observational data with that of other high-redshift blazars, we aim to construct a more comprehensive picture of blazar evolution and jet formation in the nascent universe. Understanding the accretion dynamics in these early stages is essential for a comprehensive view of SMBH growth and energy output. 
Through our detailed analysis of J1702+1301, we hope to provide valuable insights into the development of strong jets and the role of blazars in cosmic evolution, thus advancing the field of high-redshift AGN research and contributing to the broader picture of cosmic evolution.

Throughout this work, we adopt a $\rm \Lambda CDM$ cosmology with $\rm H_0 = 70 \ km \ s^{-1} Mpc^{-1}$, $\Omega_{\rm m} = 0.3$ and $\Omega_{\Lambda} = 0.7$. With this cosmological model, 1$\arcsec$ at the redshift of J1702+1301 corresponds to 5.98 kpc.

\section{observations} \label{sec:obs}
The VLBA observations of J1702+1301 were conducted in December 2021 at L, S, and C bands. The VLBA observations of J1702+1301 were conducted in December 2021 at L, S, and C bands that cover the observing frequencies from 1.35 to 7.9 GHz \citep{An2023}. At the redshift of J1702+1301, the central frequencies of these bands correspond to rest-frame frequencies of 11.3 GHz, 14.7 GHz, and 31.7 GHz, respectively, allowing us to probe the blazar’s properties at higher GHz frequencies. 

A notable challenge during these observations was the relatively low flux density state of this blazar \citep{An2023}, which impacted our ability to detect extended jet emission. This state of J1702+1301, compared to its other observing epochs, provided a unique opportunity to study the blazar's characteristics in its quiescent phase. Such observations are crucial for understanding the variability and typical emission properties of high-redshift blazars.

The data were recorded with dual polarizations and sampled at two bits, using 4 spectral windows in each band, further split into 128 channels at L and S bands or 256 baseband channels at C band with a channel width of 500 kHz. The observations were correlated at the VLBA correlator of the US National Radio Astronomy Observatory in Socorro, NM, with a correlator integration time of 2s.  Table \ref{table1} summarizes the observing parameters.

For calibration purposes, we selected J1706+1208 and J1642+3948 as fringe finders and J1707+1331 (1.3 degree away from the target, $S_{\rm S-band} \sim$ 0.3 Jy) as the phase calibrator for observations at the three frequencies. Due to the target's limited brightness, we employed phase-referencing mode with a cycle comprising 1-minute on the phase calibrator and 3.5-minute observing time on the target source. Fringe finders were observed once an hour. During the data reduction, we used J1706+1208 for fringe fitting, compensating for the unavailability of two antennas in the observations of J1642+3948. 

\begin{figure*}
   \centering
   \includegraphics[width=0.33\textwidth]{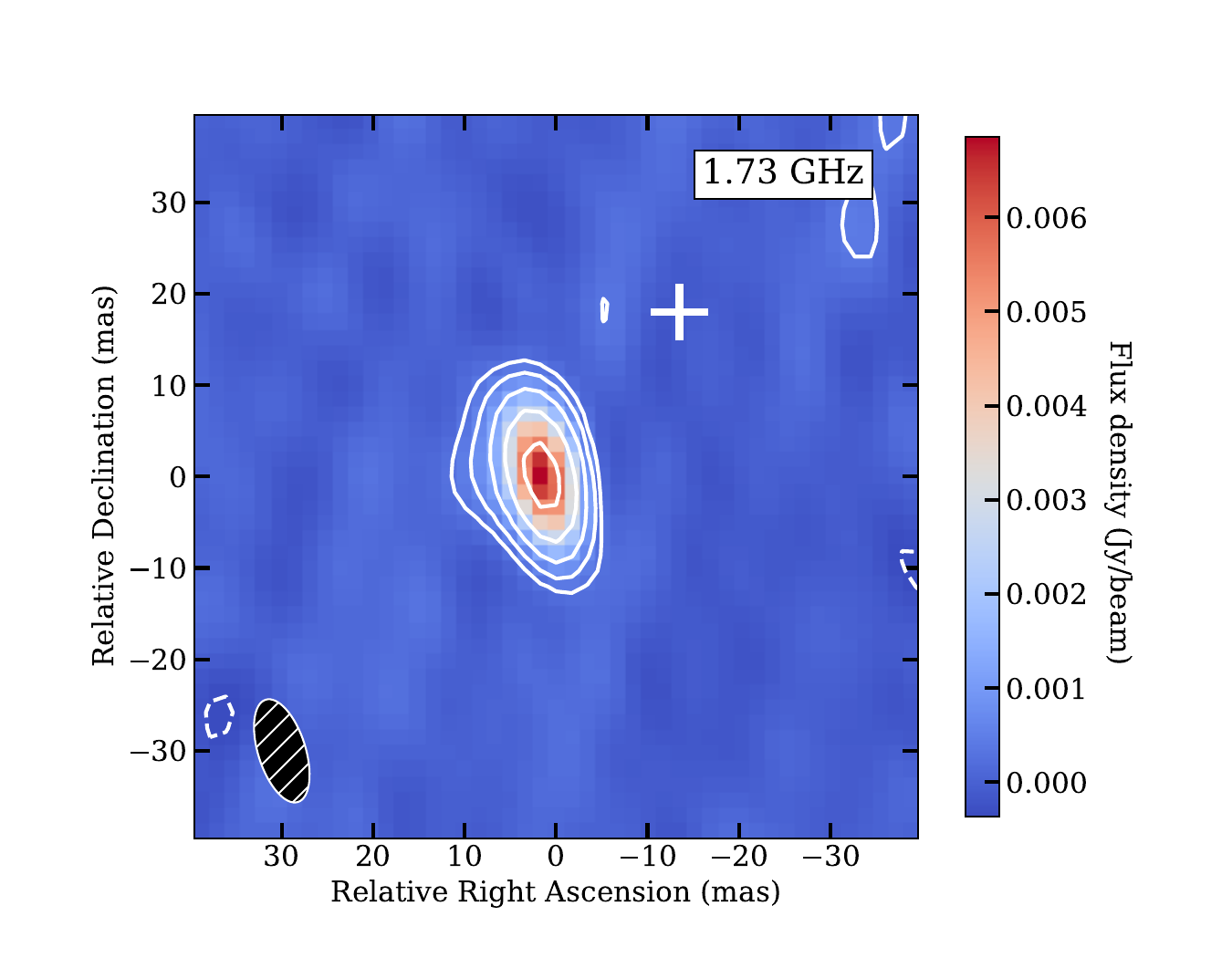}
   \includegraphics[width=0.33\textwidth]{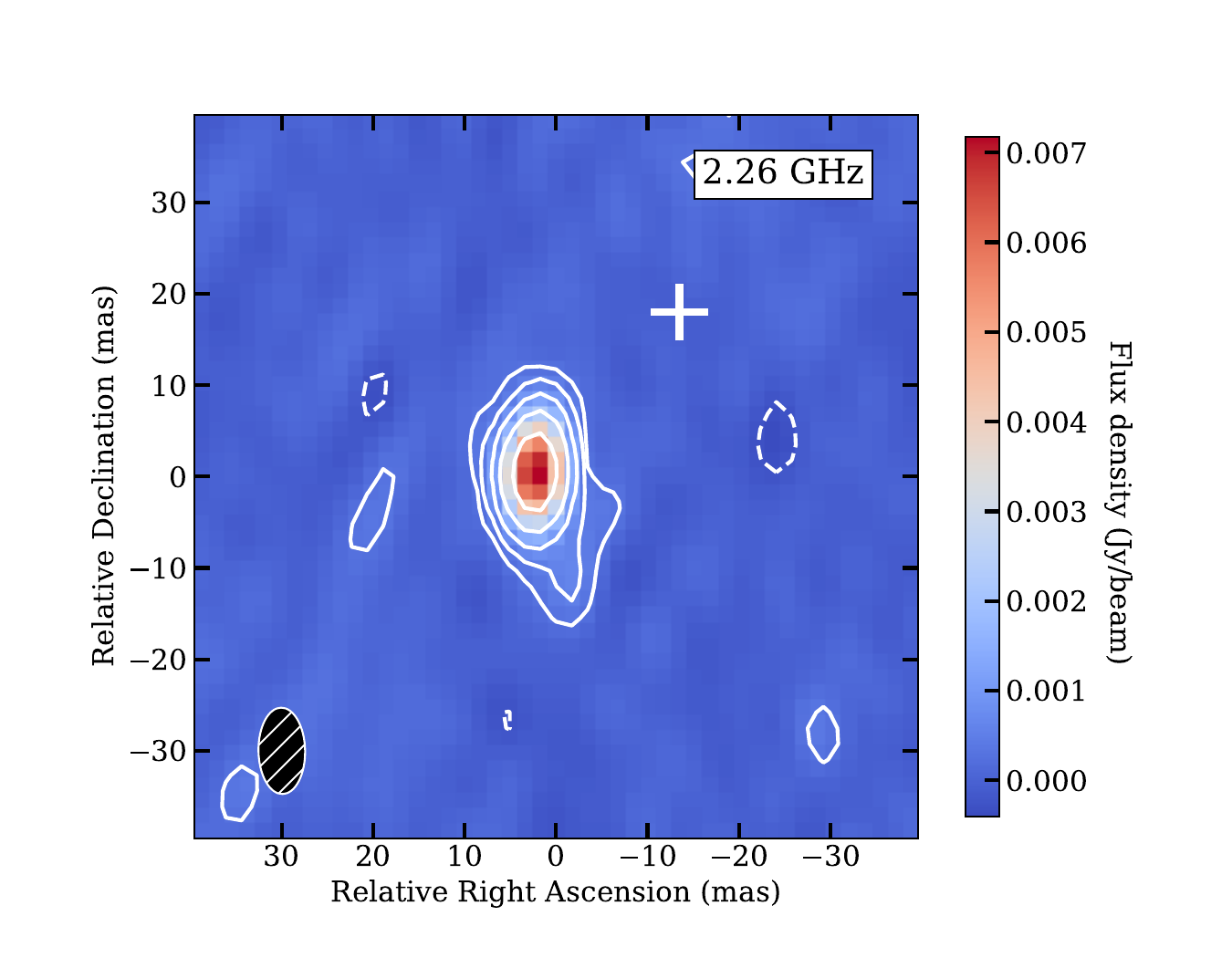}
   \includegraphics[width=0.33\textwidth]{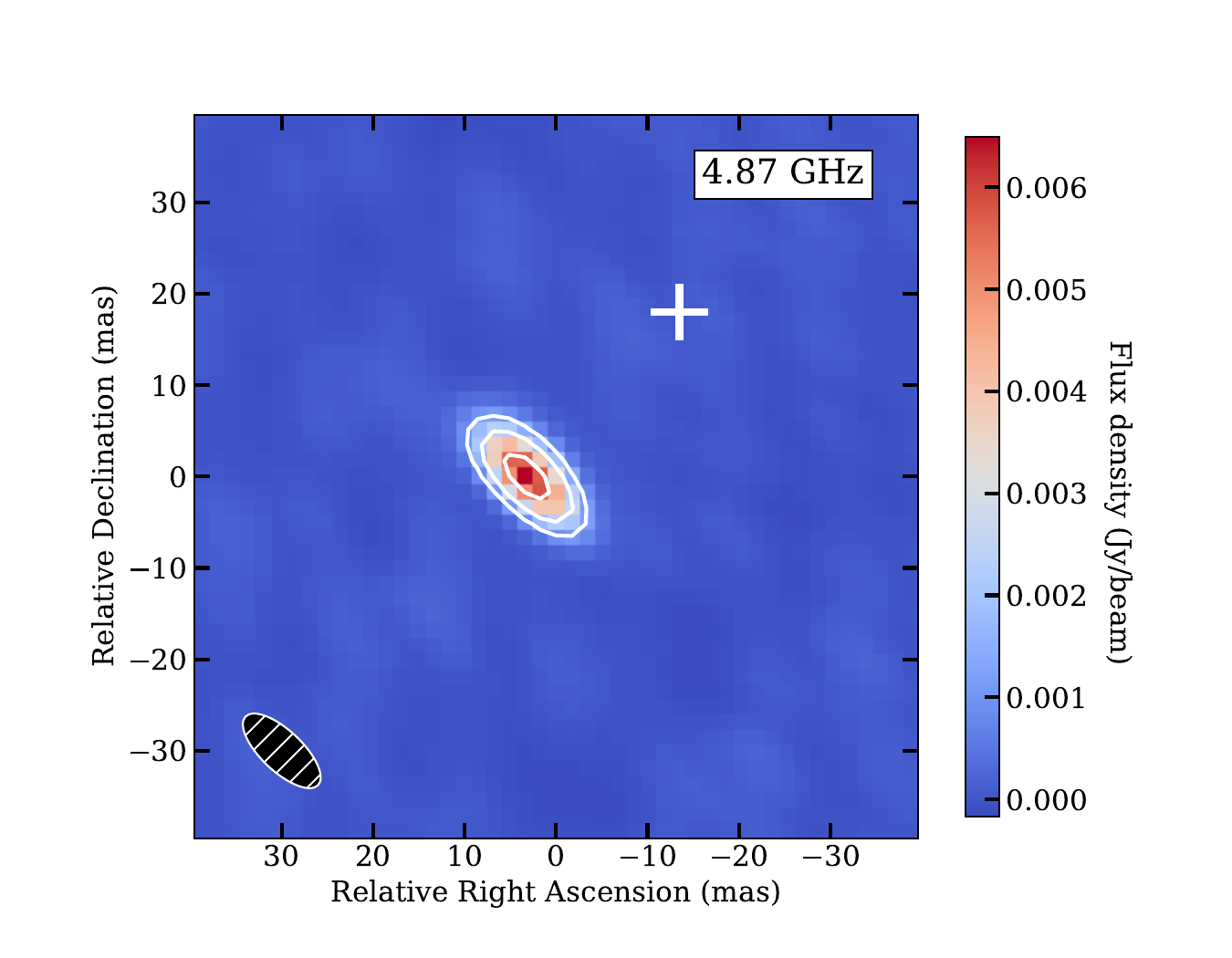}
   \caption{The VLBA radio continuum images of J1702+1301. The left, middle, and right panel shows the observations at 1.73 GHz, 2.36 GHz, and 4.87 GHz, respectively. The images are all centered at RA: 17h02m45.3119s, Dec: +13d01m02.222s, which is 22.3 mas away southeast to the measured center from optical/X-ray observations shown as a white cross. The contour levels are [-1,1,2,4,8,16,32,64]*3$\sigma$, where the $\sigma$ corresponds to the rms noise level from each image. The synthesis beam for each image is shown at the lower-left corner.
   \label{image}}
    \end{figure*}

Data reduction and analysis were performed with the Astronomical Image Processing System (AIPS; \citealt{Greisen2003}) of the NRAO. We use the ParselTongue pipeline \footnote{\url{https://github.com/SHAO-SKA/vlbi-pipeline}} deployed at the China SKA Regional Centre \citep{An2019,An2022} to automatically achieve the standard calibration steps, mainly including instrumental delay, ionospheric corrections, and fringe fitting. The pipeline also applied automatic flagging to avoid the effect of Radio Frequency Interference (RFI) and other bad data. We export the calibrated data of calibrators to the Caltech Difmap \citep{Shepherd1997} package for imaging and CLEAN task of the calibrators. Then the self-calibration for both phase and amplitude was performed on the phase calibrator based on its clean model. These self-calibration solutions were applied to the target. The overall calibration uncertainties are less than 5\%. As the final step, we applied mapping and clean procedure on the target J1702+1301 in the natural weighting. The intensity maps are shown in Fig. \ref{image}.

\begin{table*}
\centering
\caption{Observation information and image parameters.}
\begin{tabular}{cccc}\hline\hline
\textbf{Parameters}     & \textbf{L band}     & \textbf{S band}    & \textbf{C band}     \\
\hline
Observing date (in 2021)            & Dec 06  & Dec 07 & Dec 21  \\
Total observing time (hr)           & 2.75    & 4      & 2.75    \\
Central Frequency (GHz)         & 1.73    & 2.26   & 4.87    \\
Total bandwidth (MHz)   & 256     & 256    & 512     \\

Synthesis beam (mas)    & $11.73 \times 5.19$ & $9.46 \times 5.10$ & $10.91 \times 4.46$ \\
Image rms noise level (mJy beam$^{-1}$) & 0.11    & 0.12   & 0.42    \\
Peak flux density (mJy beam$^{-1}$) & 6.77    & 6.85   & 6.40    \\
Integrated flux density (mJy)  & 8.35 $\pm$ 0.09        & 7.47 $\pm$ 0.08       & 6.57 $\pm$ 0.02        \\
Size (mas)  & $3.276\pm 0.043$   & $1.626\pm 0.068$  & $0.905\pm 0.006$   \\
Brightness temperature ($10^9$K)    & 2.0     & 4.4    & 2.7   \\
\hline
Note: natural weighting is used in all the images.
\label{table1}
\end{tabular}
\end{table*}

\section{Results} \label{sec:result}

We successfully detected J1702+1301 at all three observing frequencies (1.73 GHz, 2.26 GHz, and 4.87 GHz) with remarkable signal-to-noise ratios ranging from 15 to 62. These observations were critical in determining the compact nature of the emission sources. The total intensity maps show a single, unresolved component at all frequencies (Fig. \ref{image}). This observational result is consistent across a broad field of view spanning 1--2 arcsec, with no additional features detected above a $5\sigma$ threshold ($\sigma$ being the image rms noise). The precise equatorial coordinates of the emission peak are RA: 17h02m45.3119s, dec: +13d01m02.222s, with astrometric errors of $\sim$0.1 mas arising mainly from uncertainties in the phase reference calibrator's position. The VLBI position derived at the three frequencies is consistent within the error margin. The previously reported optical/X-ray coordinates are RA: 17h02m45.3s, Dec: +13d01m02.24s, which are 22.3 mas northwest of the VLBI position. This compact VLBI component represents the jet base close to the central SMBH (see more discussion below). The radio--optical positional offset can plausibly be explained by astrometric uncertainties in the optical/X-ray measurements due to the weakness of the source in optical and X-ray images. 

We fit circular Gaussian models to the J1702+1301 VLBI source using Difmap's modelfit program. The measured total flux densities of the source were 8.35 $\pm$ 0.09 mJy, 7.47 $\pm$ 0.08 mJy, and 6.57 $\pm$ 0.02 mJy at observation frequencies of 1.73 GHz, 2.26 GHz, and 4.87 GHz, respectively (see Table \ref{table1}). The fitted Gaussian source sizes (FWHM) are $3.276\pm 0.043$ mas, $1.626\pm 0.068$ mas, and $0.905\pm 0.006$ mas at the three frequencies, respectively. At 4.87 GHz the fitted peak flux density is consistent with the integrated flux density, indicating that the emission originates from a component too compact to be resolved by the VLBA. However, at the lower frequencies of 1.73 and 2.26 GHz, our detection is limited to a single unresolved core, with integrated flux densities slightly exceeding  peak flux densities. 
The current data set, while extensive, does not allow for a conclusive determination of whether jet emission is mixed with the core emission at 1.73 and 2.26 GHz. To unravel the complexities of J1702+1301's jet morphology and its potential evolution, further investigations with higher-resolution observations are essential. 

With the measured flux density and source size, we calculated the brightness temperature, which is a crucial parameter in understanding the physical conditions of the emitting region. Employing the established formula for brightness temperature calculation from \citet{Condon1982}:
\begin{equation} \label{eq1}
T_b = 1.22\times 10^{12}\left(1+z\right) \frac{S_{\nu}}{\nu^2 \theta^2},
\end{equation}
where $\nu$ is the observing frequency in GHz, $S_\nu$ denotes the surface brightness in $\rm Jy \ beam^{-1}$, and $\theta$ represents the source size at different frequency.
We determined that the brightness temperature of J1702+1301 is $\sim 10^9$K. This measurement is particularly significant as it provides compelling evidence of the non-thermal nature of the emission. It is important to note that since the source is only marginally resolved in our observations, the fitted source sizes can be regarded as upper limits, and thus the derived brightness temperatures should be treated as lower limits. 

The high brightness temperatures, exceeding $10^9$ K at all three frequencies, provide robust evidence that the emission has a non-thermal origin. This is indicative of processes such as synchrotron emission, which are common in AGN and especially pronounced in blazars. The temperature values observed for J1702+1301, although not as extreme as some blazars in the low-redshift universe, are still remarkable for a high-redshift object and consistent well with the expected properties of blazars at such distances (for example, \citealt{Cao2014, Frey2011, Zhang2022, Spingola2020}). Note that these measurements are obtained at rest-frame frequencies of 10--30 GHz, where a general trend of decreasing brightness temperature with increasing frequency has been observed \citep{Cheng2020}. These temperatures reinforce the interpretation of J1702+1301 as a blazar \citep{Khorunzhev2021,An2023}, characterized by a highly energetic and active nucleus.

\begin{figure}
\includegraphics[width=0.5\textwidth]{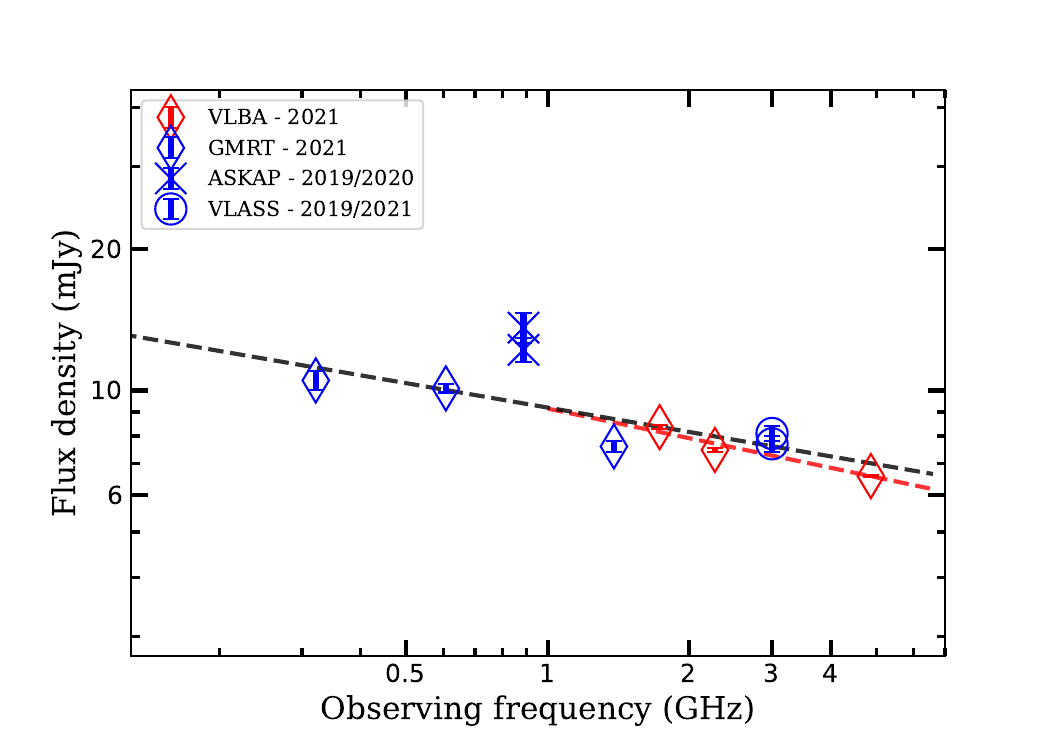}
\caption{ Radio spectral energy distribution of J1702+1301, including data from our VLBA observation (red squares) as well as archive data from the GMRT, ASKAP, and VLASS surveys (blue labels). The best fitting results of VLBA observations is shown as the red line, with a spectral index of $\alpha = -0.21\pm 0.001$. This spectrum is only slightly steeper than the previous results (black line) fitted from the synthesis imaging with much lower resolutions \citep{An2023}.
\label{index}}
\end{figure}

The spectral energy distribution (SED) analysis of J1702+1301 is crucial for understanding the radiative mechanisms and physical conditions within the blazar. 
To construct the SED, we included data from our VLBA observations (represented by red squares in Fig. \ref{index}) alongside archival data from the Giant Metrewave Radio Telescope (GMRT), the Australian Square Kilometre Array Pathfinder (ASKAP), and the Very Large Array Sky Survey (VLASS) (indicated by blue labels in Fig. \ref{index}).
Our analysis employed a standard non-thermal power-law spectrum to model the radio continuum, fitting the data across the three VLBA frequencies (corresponding to rest-frame frequencies of 9.75--33.15 GHz at the source's redshift). This approach is rooted in the form $S_{\nu} \sim \nu^{\alpha}$, where $\alpha$ is the spectral index, and $S_{\nu}$ is the flux density at frequency $\nu$. Through a least-squares fitting method, we determined the spectral index to be $\alpha = -0.21\pm 0.001$. The resulting spectrum, illustrated by the red line in Fig. \ref{index}, reveals a slightly steeper spectral index than previous results; these earlier results, represented by the black dashed line, were derived from synthesis imaging at much lower resolutions over 0.3--3 GHz \citep{An2023}. The fitted spectral index of $\alpha = -0.21$ is indicative of a flat-spectrum radio core, a characteristic commonly associated with blazars. This finding agrees well with the high brightness temperatures observed and the compact nature of the VLBI-detected radio core, providing compelling evidence of J1702+1301 being a blazar with a relativistic jet oriented close to our line of sight.


\section{discussion} \label{sec:discuss}

The VLBA observations of J1702+1301 provide valuable insights into the nature of this high-redshift AGN. The detected compact VLBI component, in close proximity to the central SMBH, likely represents the base of the jet.

The total flux densities measured at the observing frequencies (1.73 GHz, 2.26 GHz, and 4.87 GHz) were 8.35 $\pm$ 0.09 mJy, 7.47 $\pm$ 0.08 mJy, and 6.57 $\pm$ 0.02 mJy, respectively. The consistent flux densities across the observed frequencies are indicative of a compact, high-brightness temperature source, a characteristic feature of blazars. The spectral index, $\alpha = -0.21$, is consistent with expectations for a flat-spectrum radio core, reinforcing the blazar identification of J1702+1301.

One notable aspect of our study is the positional offset between the radio and optical/X-ray coordinates of J1702+1301. The positional offset of 22.3 mas between the radio and optical/X-ray coordinates is significant (see Section 3). While astrometric uncertainties in the optical and X-ray bands could partly account for this discrepancy, particularly given the source's relative weakness in these bands, this substantial offset is uncommon in blazar observations. Radio-optical offsets in other blazars are usually much smaller. This discrepancy may indicate more complex jet dynamics than typically observed or could suggest variations in the regions of emission. 

The nature of the compact VLBI component, especially its high brightness temperature and flat radio spectrum, reinforce the significance of J1702+1301 as a blazar. This conclusion is consistent with the observed characteristics of other high-redshift blazars and contributes to our broader understanding of AGN jet formation and evolution in the early universe. 

The RATAN-600 telescope has been monitoring J1702+1301 since late 2020 at 4.7, 8.2, and 11.2 GHz as shown in Fig.5 in \citealt{An2023}. 
According to the flux density curve, we note that our VLBA observations were carried out when J1702+1301 was in a relatively low flux density state compared to the active state after February 2022. This timing likely explains the absence of extended jet emission in our VLBI images. Such a state provides a baseline for the blazar's `normal' condition. Given J1702+1301's high brightness temperature and rapid variability, upcoming VLBI observations during its flaring state are highly anticipated. Comparing these future observations with our current data will enhance our understanding of blazar behavior across different activity phases, especially for high-redshift blazars.

Similar to the local Universe, only a small fraction ($\sim 10 \%$) of high-$z$ AGN have large radio loudness \citep{Kratzer2015, Liu2021}. In particular, a limited number of radio-loud AGN with radio loudness $R>$10 have been identified at $z>$5. The properties of J1702+1301 are consistent with those observed in other high-redshift blazars, such as J0309+2717 (\citealt{Belladitta2020,Spingola2020}, $z = 6.10$), Q0906+6930 (\citealt{Romani2004}, $z=5.47$), and SDSS J102623.61+254259.5 (\citealt{Sbarrato2012},$z=5.3$). These blazars exhibit strong radio and X-ray emission, flat radio spectra, and radio variability -- typical features of blazars. In particular, both J0309+2717 and Q0906+6930, observed with VLBI, show less powerful jets with lower proper motion and bulk Lorentz factors compared to typical low-redshift blazars. However, the high brightness temperature and rapid variability observed in J1702+1301 suggest that it could be one of the most active blazars at redshifts $> 5.5$. This is in contrast to the less powerful jets observed in the above blazars, suggesting a potentially unique jet behavior in J1702+1301. The unique characteristics of J1702+1301, compared to other high-redshift blazars, manifests the diversity in blazar properties and behaviors, suggesting a more complex picture of AGN activity in the early stages of the Universe.


\section{summary} \label{sec:summary}

The study of the SRGE J170245.3+130104, located at a redshift of $\sim$5.5, represents a significant step in understanding high-redshift AGN and their jet formations. Our observations with the VLBA have successfully detected J1702+1301 across multiple frequencies, yet its jet structure remains unresolved, highlighting the challenges in studying such distant objects.

The VLBA observations at 1.73 GHz, 2.26 GHz, and 4.87 GHz reveal a compact, unresolved core, indicative of a highly active AGN. The spectral index derived from these observations, $\alpha = -0.21$, is consistent with previous low-resolution studies and supports the interpretation of a flat-spectrum radio core. This finding, coupled with the high brightness temperatures observed, reinforces J1702+1301's classification as a blazar, particularly notable for its rapid variability and exceptional brightness at such a high redshift.
Our findings, particularly the observations conducted during the blazar's low flux density state, provide a valuable baseline for understanding its typical emission state. This contrasts with potential future observations during higher activity periods, offering a dynamic view of the blazar's radio variability. 

J1702+1301 contributes to the limited but growing sample of blazars known at redshifts greater than 5, providing unique insights into the early stages of jet formation and evolution in young AGN. Our study highlights the importance of further high-resolution observations at varying frequencies to fully understand the jet dynamics and emission characteristics of J1702+1301. Such future observations are expected to offer more detailed information on the jet properties, especially during periods of flaring activity.
J1702+1301 shows unique characteristics of rapid variability when compared to other $z>5$ blazars.  This comparative analysis with other distant blazars indicates the heterogeneity in their emission properties, challenging and enriching our current understanding of blazars in the early Universe.

\begin{acknowledgements}
This research has been supported by the National SKA Program of China (2022SKA0120102, 2022SKA0130103). 
We gratefully acknowledge financial support from Pinghu Laboratory.  
This project is funded from China Postdoctoral Science Foundation (certification number: 2023M733625).
YQL is supported by the Shanghai Post-doctoral Excellence Program and Shanghai Sailing Program under grant number 23YF1455700. 
SGG is supported by Youth Innovation Promotion Association CAS Program under NO.2021258. 
ALW is thankful for the financial support received from the University of Chinese Academy of Sciences, appreciates the support and hospitality provide by the SKA Observatory and Jodrell Bank Centre for Astrophysics at the University of Manchester.
This work used resources of China SKA Regional Centre prototype funded by the Ministry of Science and Technology of the People’s Republic of China and the Chinese Academy of Sciences.
The National Radio Astronomy Observatory (NRAO) is a facility of the National Science Foundation operated under cooperative agreement by Associated Universities, Inc. This paper makes use of the VLBA data from program BR240.

\end{acknowledgements}

%
%

\bibliography{aa}{}
\bibliographystyle{aasjournal}

\end{document}